\documentclass[aps,prl,twocolumn,groupedaddress,showpacs,tighten,nofootinbib]{revtex4}
\usepackage{graphicx}
\usepackage{times}
\def\beq{\begin{equation}}
\def\eeq{\end{equation}}
\def\bey{\begin{eqnarray}}
\def\eey{\end{eqnarray}}

\def\lsim{\mathrel{\raise.3ex\hbox{$<$\kern-.75em\lower1ex\hbox{$\sim$}}}}
\def\gsim{\mathrel{\raise.3ex\hbox{$>$\kern-.75em\lower1ex\hbox{$\sim$}}}}

\begin{document}

\title{What Can Gamma Ray Bursts Teach Us About Dark Energy?}
\author{Dan Hooper$^1$ and Scott Dodelson$^{1,2}$}
\affiliation{$^1$Fermi National Accelerator Laboratory, Particle Astrophysics Center, Batavia, IL  60510-0500}
\affiliation{$^2$Department of Astronomy \& Astrophysics, The University of Chicago,
Chicago, IL~~60637-1433}

\date{\today}

\begin{abstract}

It has been suggested that Gamma Ray Bursts (GRB) may enable the expansion rate of our Universe to be measured 
out to very high redshifts ($z \gsim 5$) just as type Ia supernovae have done at $z \sim$1--1.5. 
We explore this possibility here, and find that GRB have the potential to detect dark energy at high statistical
significance, but they are unlikely to be competitive with future supernovae missions, such as SNAP, in measuring the
properties of the dark energy. The exception to this conclusion is if there is appreciable dark energy at early times, in which 
case the information from GRB's will provide an excellent complement to the $z\sim 1$ information from supernovae.
\end{abstract}
\pacs{98.70.Rz; 95.36.+x
\hspace{0.5cm} FERMILAB-PUB-05-532-A}
\maketitle

\section{Introduction}

By observing distant type Ia supernovae (SNe Ia), the expansion history of our Universe has been studied out to redshift of order unity. 
These observations have showed that the expansion rate of our Universe is accelerating \cite{accelerating}, which is generally attributed 
to the presence of dark energy. These observations allow us to measure the expansion rate only over recent times, however. 
If our Universe's expansion rate could be measured over a range of higher redshifts, that could provide a useful discriminator 
of various models of dark energy -- especially those models with a sizable quantity of dark energy at early times.

With this in mind, we here consider using Gamma Ray Bursts (GRB) as cosmological probes. The primary advantage of GRB over SNe Ia 
is that they are considerably brighter and thus can be observed at much higher redshifts. The {\it average} GRB observed by the 
currently operating Swift satellite has a redshift of $\sim2.8$ \cite{swiftmean}, whereas the highest redshift SNe Ia among the 
157 in the "gold" sample of Ref.~\cite{gold} is less than 1.8. Swift is expected to detect several GRB each year at $z > 5$. 

GRB have substantial disadvantages in comparison to Ia SNe, as well. SNe Ia are powerful probes of our Universe's expansion 
because their intrinsic luminosity can be inferred independently of any redshift measurement -- ie. they are standardizable candles. 
The degree to which GRB can be used as standard candles is not yet fully known. Although the quantity of (geometrically corrected) 
energy released in the jets of long duration GRB is clustered around $1.3\times 10^{51}$ ergs \cite{ejet}, the dispersion of this 
quantity is too large to be of much use in studying cosmology. This dispersion can be considerably reduced, however, by making use of 
the various known correlations between the luminosity of a burst and its other observable parameters. Such correlations include those 
associated with the variability \cite{variability} and spectral lag \cite{lag} of a GRB. 
In addition to these, two other well-known GRB luminosity indicators have appeared in the literature \cite{other}. 
We will not discuss these here, as they are either too inaccurate or poorly established to be adopted reliably at this time. 
It appears at least plausible, however, that observations by Swift and other experiments will improve our ability to accurately determine 
the intrinsic luminosity of GRB. The degree to which this can be accomplished will determine the usefulness of GRB to cosmology.

Here we study how much can be learned about cosmological expansion from GRB's. In particular, we ask whether future Swift observations will
pin down properties of the dark energy, and if so, how these projected constraints compare with the expected results from Ia SNe.

\section{Standard Dark Energy Parameters}

By measuring the apparent magnitude $m$ of a standard candle with known absolute magnitude $M$, we can infer the distance modulus at the redshift of interest:
\beq
\mu(z) \equiv m-M = 5\log_{10}\left( {d_L(z) \over 10\, {\rm pc}} \right)
.\eeq
The luminosity distance is an integral over the time varying Hubble rate:
\beq
d_L(z) = (1+z) \int_0^z {dz'\over H(z')}
,\label{dl}\eeq
so, in principle, we can hope to learn about the expansion history of the universe by measuring distance moduli at a
variety of redshifts.

With this mind, consider a popular model for the expansion history:
\beq
H(z) = H_0 \left[ \Omega_m (1+z)^3 + (1-\Omega_m) (1+z)^{3(1+w)} \right]^{1/2}.
\label{hub}\eeq
Here $H_0=100h$ km sec$^{-1}$ Mpc$^{-1}$ is the current value of the Hubble constant; $\Omega_m$ is the matter
density in units of the critical density; and $w$ is the equation of state of the dark energy, assumed constant. This model assumes
that the universe is flat so that $\Omega_{\rm DE} = 1-\Omega_m$. We assume throughout that the true model has $h=0.7$ and $\Omega_{\rm DE}=0.7$.

\begin{figure}
\resizebox{8cm}{!}{\includegraphics{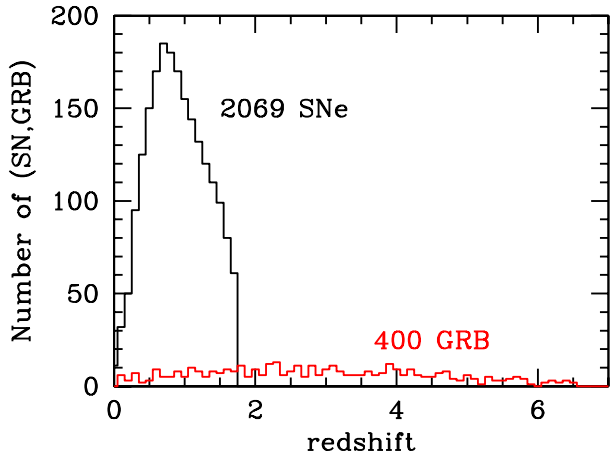}} 
\caption{Redshift distributions of Ia Sne and GRB from simulations of future SNAP and Swift missions.}
\label{histo}
\end{figure}

Let us ask how future observations of Ia SNe and GRB will constrain these parameters. We simulate a sample of 2069 Ia SNe 
with the redshift distribution expected in the SNAP mission~\cite{snapdist} and an intrinsic dispersion 
in the distance modulus of $\sigma = 0.16$. The currently operating Swift satellite is able to detect roughly $\sim$100 long-duration GRB per year, about half 
of which include a redshift measurement. We assume 400 GRBs with redshifts drawn from a redshift 
distribution following the star formation rate (which has been shown to be consistent with the first Swift data~\cite{swiftmean}). 
The redshifts from these simulations are shown in Fig.~\ref{histo}. 

Besides the smaller numbers, GRB suffer from a larger internal dispersion.
The scatter observed in measurements of variability and spectral lag in GRB are roughly 
$\sigma_{\log({\rm{lag}})} \sim$ 0.3--0.4 and $\sigma_{\log({\rm{var}})} \sim 0.2$. This 
ultimately leads to a dispersion in GRB magnitude of roughly 0.6 to 0.7, which is about four times 
worse than with Ia SNe \cite{hubble}. We use $\sigma=0.64$ as an estimate of the current internal dispersion of GRBs, but 
argue that this is conservative.
Liang and Zhang \cite{other} have considered a wider range of GRB observables to construct an empirical relationship to determine GRB 
luminosity. Using this technique, they reduce the dispersion to $\sigma \sim 0.45$. We find it likely that future observations of large 
numbers of GRB will reveal stronger luminosity indicators, which translate to even smaller dispersions. 

\begin{figure}
\resizebox{8cm}{!}{\includegraphics{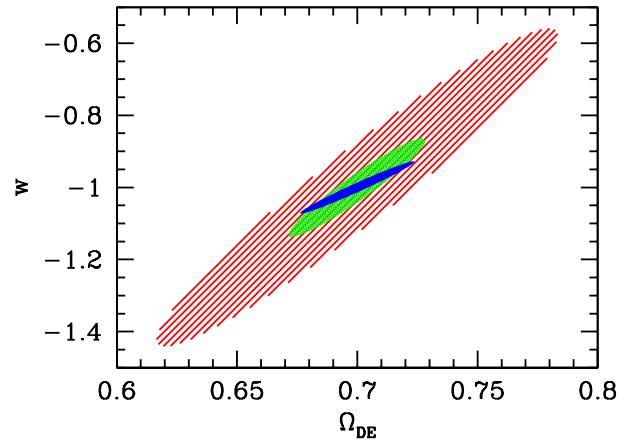}} 
\caption{Constraints on dark energy parameters that could be obtained from 2069 Ia SNe (e.g., SNAP) and 400 GRB (e.g., Swift).
The weakest contsraints (red hatched) come from GRB with the internal dispersion 
equal to the current (conservative) value of $0.64$. If correlations can be found which reduce the
internal dispersion to $0.16$ (green, parallel to other GRB), then the constraints become tighter, but still looser than 
those from Ia SNe with dispersion $0.16$ (innermost blue). Part of the reason that the GRB constraints are weaker is that
lensing by large scale structure is more damaging for high redshift sources.}
\label{ellow}
\end{figure}

Fig.~\ref{ellow} shows the projected constraints from SNAP and Swift on dark energy parameters. Even if the internal dispersion of GRBs gets no better than
its present value, Swift observations should be able to detect dark energy at high statistical significance. That is, even the outermost contours in
Fig.~\ref{ellow} represent a $>5\sigma$ detection of dark energy. This will be an important confirmation of the SNe results because it is possible, 
although unlikely, that the SNe observations are not the result of accelerated expansion, but rather of 
other effects, such as the absorption of the light from Ia SNe by dust \cite{dust}. Although this hypothesis has 
been challenged~\cite{dustproblems}, it will be nice to have independent confirmation.\footnote{Combined measurements of the CMB and
matter density suggest that the total density is more than three times larger than the matter density. This is another strong argument for
dark energy.} At the energies at which GRB are 
observed (keV-MeV), the effects of dust extinction should be entirely negligible. Therefore a measurement of our 
Universe's expansion history using GRB could exclude the possibility of dust mimicking the effects of accelerated 
expansion in the observations of Ia SNe.\footnote{It has also been suggested that the dimming of Ia SNe might be 
the effect of the oscillating of photons into axions instead of accelerated expansion \cite{axion}. This occurs 
for both optical and gamma-ray photons, however, and therefore cannot be distinguished using GRB.} 

While Swift observations could provide independent evidence for dark energy, Fig.~\ref{ellow} shows that the anticipated statistical constraints on standard
dark energy
parameters are not competitive with those expected from Ia SNe. Even if the GRB internal dispersion can be reduced, the constraints from GRB will probably not
be as strong as those from SNe. The green contour in Fig.~\ref{ellow}
shows that, even if the internal dispersion is reduced by a factor of $4$, the GRB sample will be less constraining than SNe. Part of this conclusion is based
on the effects of gravitational lensing~\cite{lensing}. The additional dispersion due to the inevitable lensing by large scale structure 
increases with source redshift. For SNe, the
lensing dispersion is likely to be smaller than the internal dispersion, but for GRB's at high redshift, lensing significantly loosens constraints. For exmaple, in our
best-case scenario, wherein the internal dispersion is $0.16$, the dispersion due to lensing becomes greater than the internal dispersion at redshift $2.7$.

%

We note that other authors have estimated the ability of GRB to constraint cosmological parameters, 
including $\Omega_{\Lambda}$ \cite{other,hubble,other2}. Although our results appear to be consistent 
with these studies, we emphasize that any conclusions depend critically on the dispersion assumed in any future GRB sample.

\section{Model-independent Constraints on the Expansion History}

Conclusions about the relative merits of SNe and GRB's depend on the underlying model.
We are led to ask therefore how well each will do in constraining the expansion history in a model-independent way.
As demonstrated by Wang and Tegmark~\cite{wang}, distance modulus data can be used to constrain $H(z)$ without assuming any underlying model.
Eq.~\ref{dl} can be discretized to read
\beq
d_{L,i} = A_{ij} H^{-1}_j
\eeq
where the kernel $A_{ij}=(1+z_i)\Delta z$ and we have assumed the Hubble radius is constant over bins of width $\Delta z$. 
Written this way, it is clear that extracting the expansion history $H^{-1}_j$ or equivalently $H(z)$ is tantamount to 
solving an inversion problem, a problem that cosmologists have dealt with successfully in several contexts~\cite{book}. 
Here we follow the technique of Wang and Tegmark~\cite{wang} to project uncorrelated error bars on the expansion
history from current and future surveys.

\begin{figure}

\resizebox{8cm}{!}{\includegraphics{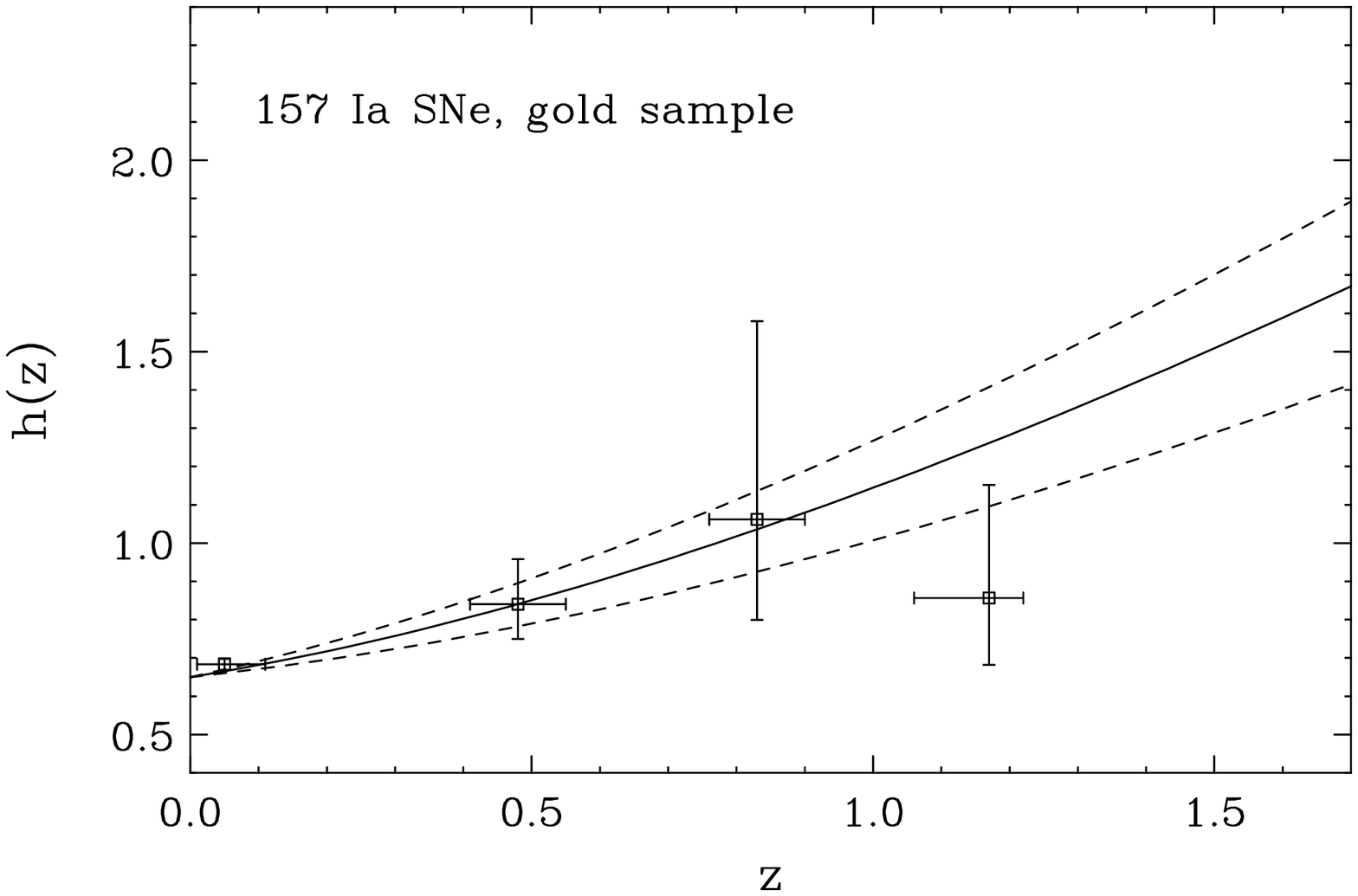}} \\
\resizebox{8cm}{!}{\includegraphics{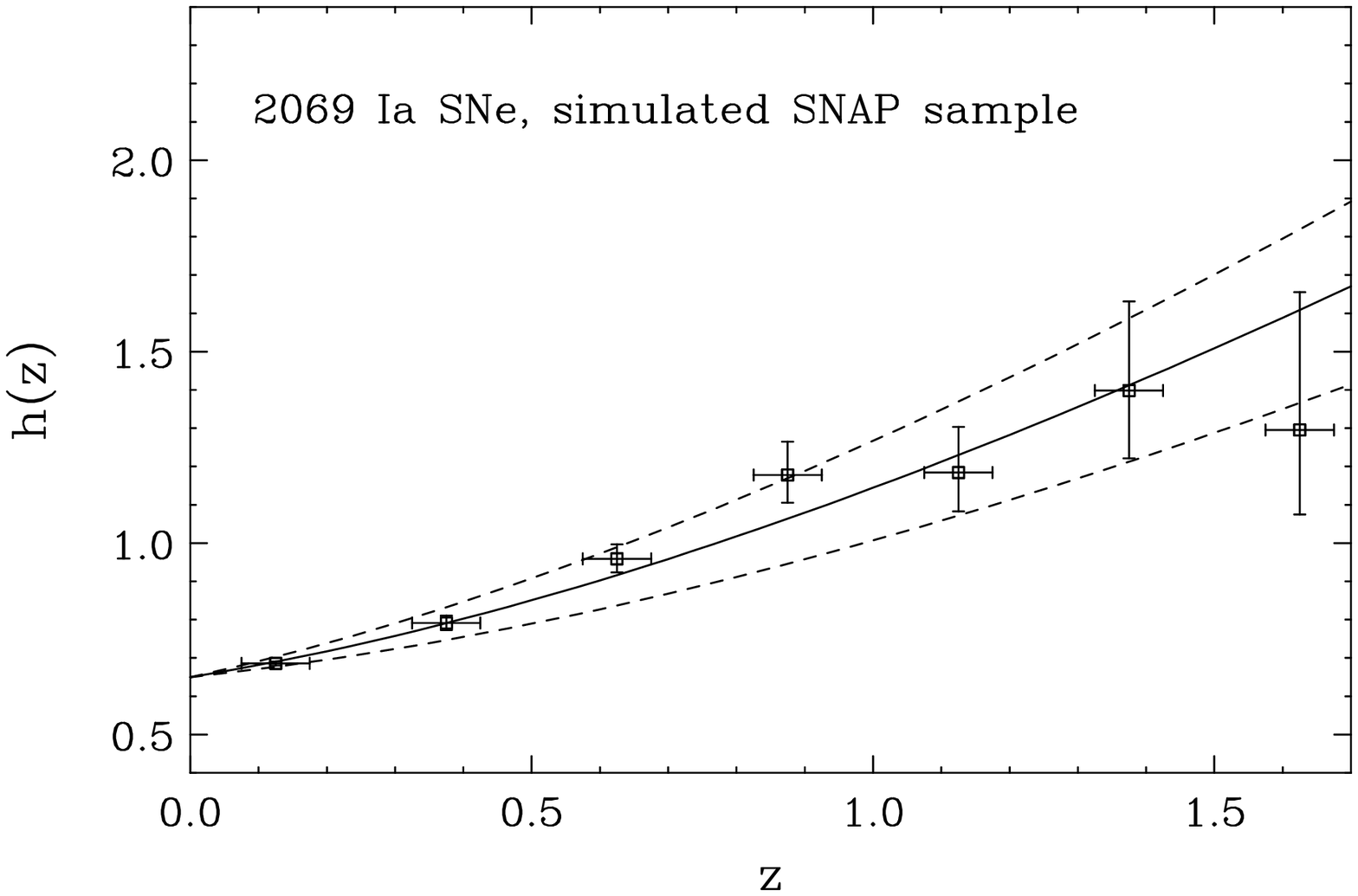}}

\caption{The ability of Ia SNe to constrain the expansion history of our Universe. 
In the top frame, results from the 157 Ia SNe of the "gold" sample \cite{gold} are shown. 
In the bottom frame, the results from a simulated sample of 2069 Ia SNe with $\sigma=0.16$ are shown, 
as is expected for a future SNAP mission. The solid line represent a model 
with $\Omega_{\Lambda}=0.7$, $h(0)=0.65$ and a flat Universe. The dashed lines are similar models with 
$\Omega_{\Lambda}=0.6$ and 0.8. In the case of the simulated data, the $\Omega_{\Lambda}=0.7$ model is the "true" model.}
\label{sn}
\end{figure}

Our ability to constrain the expansion history of our Universe using Ia SNe is shown in figure~\ref{sn}. In the top frame, 
the status of the current data consisting of 157 Ia SNe (the "gold" sample) is shown. In the bottom frame are results for 
a simulated sample of Ia SNe corresponding to the expectations of a future SNAP mission. 
Figure~\ref{sn} demonstrates that Ia SNe are effective tools for measuring the expansion rate of our Universe in a model-independent
way. The current 
data provide some degree of information out to $z ~$1--1.5, and SNAP should be able to map this history out to $z\sim 1.7$ (the maximum
redshift in the simulation) with some 
accuracy. Nothing can be revealed at higher redshift using these objects, however. 

The range of redshifts which can be explored using Ia SNe is the range in which dark energy is expected to play an important 
role in most models. For example, if the dark energy of our Universe is a cosmological constant, then the ratio of the 
matter density to the density of dark energy is 
$\rho_{m}/\rho_{\Lambda} = (1+z)^3 \Omega_m /(1-\Omega_m)$. If this is the case, then 
by $z=1.5$, $\rho_{m}/\rho_{\Lambda} \sim 7$ and dark energy has little effect at redshifts larger than those studied by Ia SNe. 
In this, as in most models, the Universe was highly matter dominated with little in the way of dark 
energy at $z>1.5$. It is important to note, however, that we do not yet have any observational evidence that these models are correct. Our understanding 
of our Universe's evolution is based upon the observations of three times in its history: Big Bang Nucleonsyntheis (BBN), the formation 
of the Cosmic Microwave Background (CMB) and the recent epoch, including Ia SNe measurements. The period between $z\sim1000$ and $z\sim2$ is, 
therefore, very poorly constrained.

\begin{figure}

\resizebox{8cm}{!}{\includegraphics{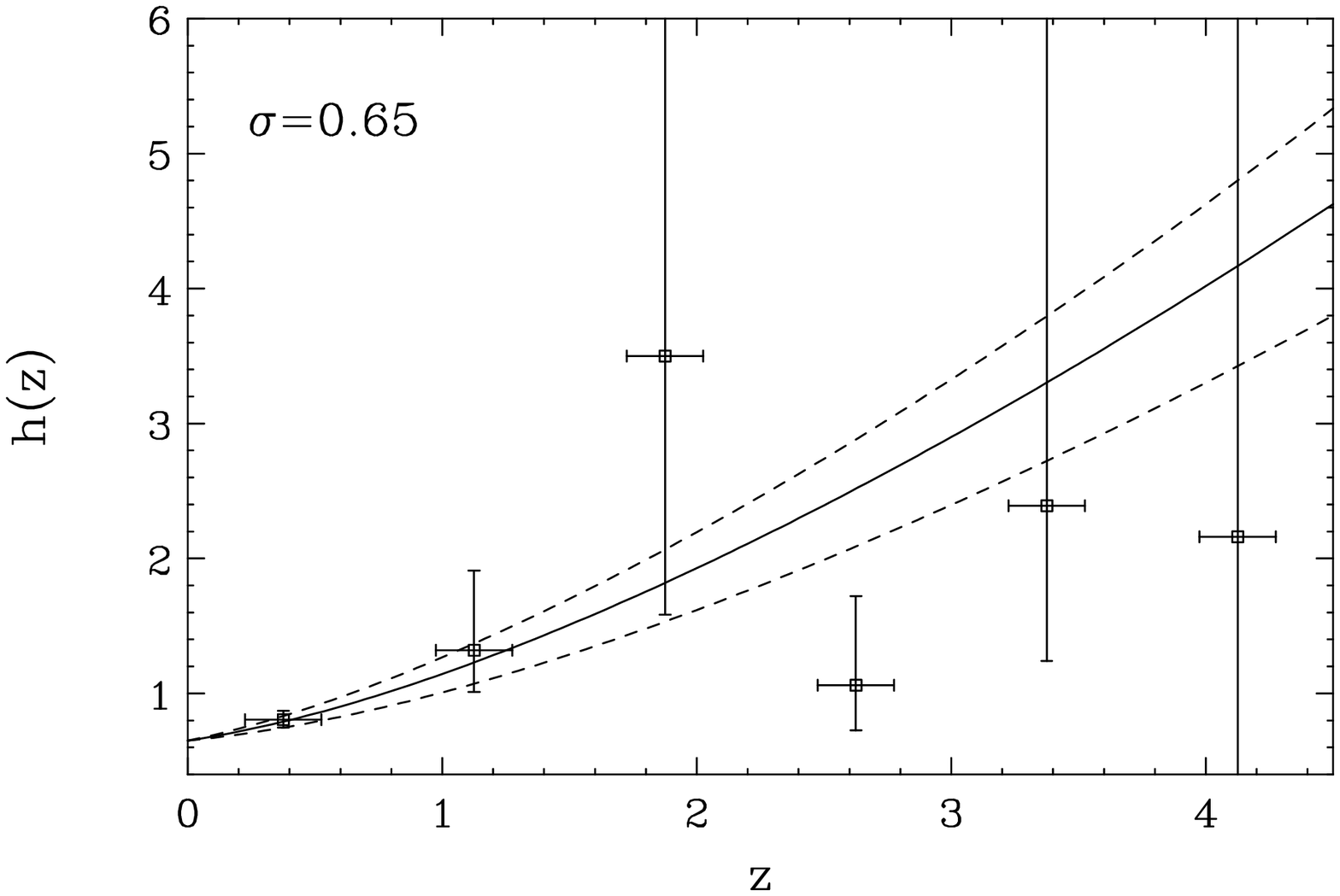}} \\
\resizebox{8cm}{!}{\includegraphics{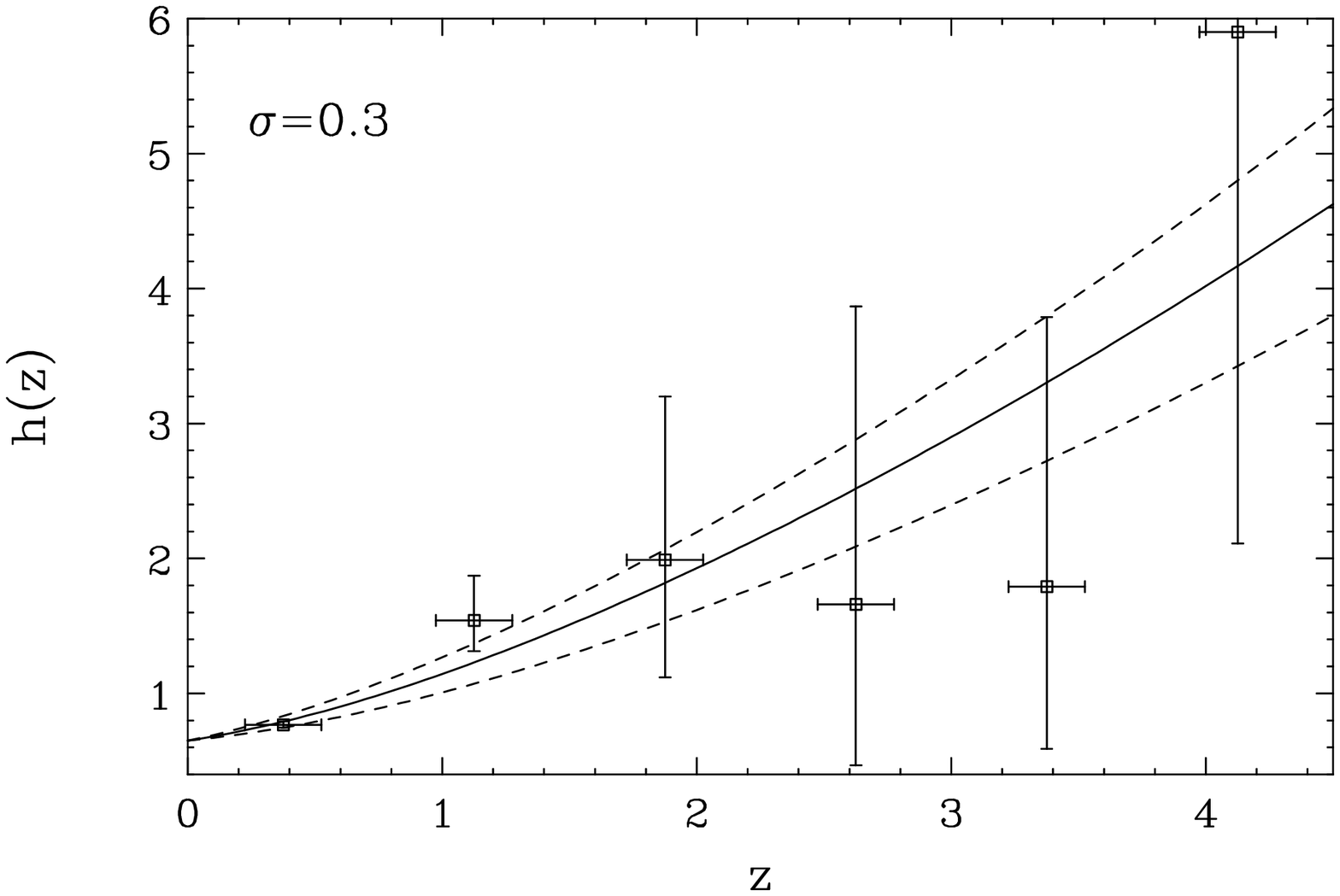}} \\
\resizebox{8cm}{!}{\includegraphics{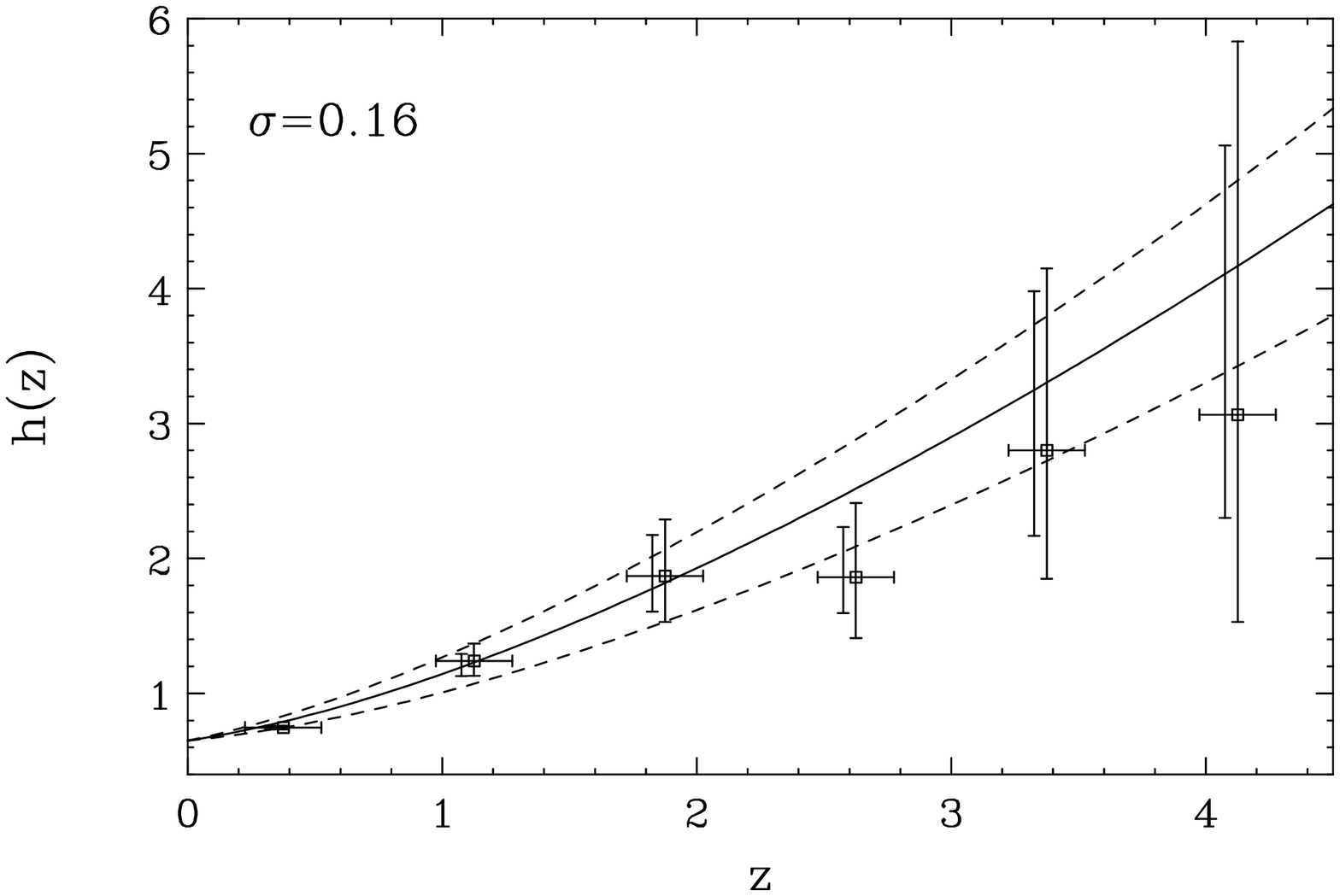}}
\caption{The ability of 400 Swift GRB's to constrain the expansion history of our Universe in a model-independent fashion. 
From top to bottom, panels show GRB constraints
on the expansion history as intrinsic dispersion is lowered from $0.65$ to $0.30$ to $0.16$. In all panels, curves are $\Lambda$ models with 
$\Omega_\Lambda=0.7$ (solid), $0.6$ (upper dashed), $0.8$ (lower dashed). In the lowest frame, also shown are error bars without the additional dispersion due to gravitational lensing (slightly offset to the left).}
\label{grb}
\end{figure}

GRB, which can be observed at higher redshift, will be able to constrain the expansion history of our Universe at earlier times than SNe.
This is illustrated in figure~\ref{grb} for samples of 400 Swift GRB with an intrinsic dispersion in their brightness of $\sigma=$0.65, 0.30 and 0.16. 
Consider the (optimistic) lower panel, for which the internal dispersion is 0.16. In that case the Hubble rate at redshift 2 would be constrained to within $\sim20\%$.
This means that the dark energy at redshift 2, $\propto (H/H_0)^2 - 27\Omega_m$, would be constrained to within $\sim 50\%$ since other experiments will likely 
pin down the matter
density to within a percent or two. Such a constraint would be an important
bridge in the gap between the constraints at redshift unity and those at $z\sim 1000$.
In short, cosmological measurements with GRB can be quite a powerful 
tool, extending the Hubble diagram to much higher redshifts than Ia SNe.


\section{Parametrized Early Dark Energy Model}

To drive home the point that GRB are powerful probes of early dark energy models, 
we focus on one model in which a non-negligible quantity of dark energy is present at high redshifts~\cite{quigg1,quigg2}. In particular, 
we consider a model with an oscillating equation of state
\begin{equation}
w(a) = -\cos(B~\ln a),
\label{olga}
\end{equation}
where $a$ is the scale factor and $B$ is a free parameter. In this model,  the equation of state of the dark energy currently is
close to -1, as is observed, but varied between 1 and -1 in 
past times. After radiation domination, the Universe underwent periodic eras of matter and dark energy domination, 
thus providing an explanation for the "why now" problem \cite{quigg1}. The rate of these oscillations is consistent with current observations over the range 
$0.4 \lsim B \lsim 2.0$ \cite{quigg2}. 

\begin{figure}
\resizebox{8cm}{!}{\includegraphics{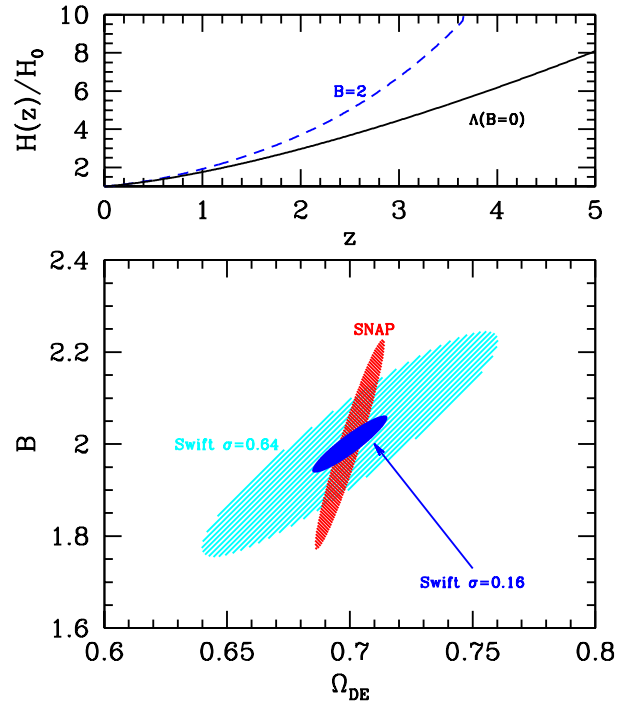}}
\caption{Projected constraints on the early dark energy model of Eq.~(\ref{olga}) with $B=2$. Top panel shows the evolution
of the Hubble rate in the model as compared with the evolution in $\Lambda$CDM (which corresponds to $B=0$.}
\label{ellob}
\end{figure}

If the of value of $B$ is sufficiently large, then dark energy would be prevalent at redshifts measurable with GRB. The ability of Swift to constrain 
such a scenario is shown in figure~\ref{ellob}. Here the complementarity of SNe and GRB measurements is striking. The parameter $B$ dictates
the level of dark energy at high redshifts where SNe have little constraining power. With small internal dispersion GRB can measure the dark energy abundance at
high redshift ($\sim 3$) and therefore place tight constraints on $B$. Even with the current value of the internal dispersion, 400 GRB's would constrain
$B$ as tightly as SNe.

\section{Conclusions}

In this article, we have studied the possibility of constraining the properties of dark energy using future observations of 
Gamma Ray Bursts (GRB). GRB are brighter than Supernovae (SNe), and thus can be observed at much higher redshifts. The Swift satellite, 
for example, can measure several GRB at redshift above 5 each year. This is in contrast to type Ia SNe observations, which are limited to redshifts smaller than 1.7, even with an experiment such as SNAP. 

The weakness of using GRB as cosmological probes lies in the degree of scatter in their intrinsic luminosities. 
The degree to which GRB are standardizable candles is not yet well known. Currently known relationships between 
GRB luminosity and other independent observables (such as variability and spectral lag) have led to magnitude 
dispersions on the scale of $\sigma \sim 0.6$. With this level of dispersion, GRB should be able to detect 
dark energy at a significance above 5$\sigma$, thus providing an independent verification of SNe observations.

It is quite plausible, however, that future observations will enable 
GRB luminosites to be determined with substantially higher accuracy. Even if the magnitude dispersion of GRB can 
be substantially reduced, however, these objects will probably never compete with SNe as probes of dark energy. Part of the reason
for this is that lensing by large scale structure becomes more important for sources at high redshifts. This results in a floor on the
total dispersion of high redshift GRB and ultimately on the statistical error on the dark energy equation of state.

The one caveat to this conclusion is if there is appreciable early ($z>1.5$) dark energy. In that case,
GRB will provide a useful and complementary probe of the expansion history of our Universe.

{\it Acknowledgments:} We would like to thank Chris Quigg, Mike Stamatikos and Gajus Miknaitis for helpful discussions. 
This work has been supported by the US Department of Energy and by NASA grant NAG5-10842.


\begin{thebibliography}{}

\bibitem{accelerating}
 A.~G.~Riess {\it et al.}  [Supernova Search Team Collaboration],
  Astron.\ J.\  {\bf 116}, 1009 (1998)
  [arXiv:astro-ph/9805201];
  S.~Perlmutter {\it et al.}  [Supernova Cosmology Project Collaboration],
  Astrophys.\ J.\  {\bf 517}, 565 (1999)
  [arXiv:astro-ph/9812133].



\bibitem{swiftmean}
  P.~Jakobsson {\it et al.},
  arXiv:astro-ph/0509888.

\bibitem{gold}
  A.~G.~Riess {\it et al.}  [Supernova Search Team Collaboration],
  Astrophys.\ J.\  {\bf 607}, 665 (2004)
  [arXiv:astro-ph/0402512].

\bibitem{ejet}
 J.~S.~Bloom, D.~A.~Frail and S.~R.~Kulkarni,
  Astrophys.\ J.\  {\bf 594}, 674 (2003)
  [arXiv:astro-ph/0302210];
  D.~A.~Frail {\it et al.},
  Astrophys.\ J.\  {\bf 562}, L55 (2001)
  [arXiv:astro-ph/0102282].

\bibitem{variability}
  E.~E.~Fenimore and E.~Ramirez-Ruiz,
  arXiv:astro-ph/0004176;
  D.~E.~Reichart {\it et al.},
  Astrophys.\ J.\  {\bf 552}, 57 (2001);
  D.~E.~Reichart and M.~C.~Nysewander,
  arXiv:astro-ph/0508111.

\bibitem{lag}
  J.~P.~Norris, G.~F.~Marani and J.~T.~Bonnell,
 Astrophys.\ J.\  {\bf 534}, 248 (2000);
J.~D.~Salmonson, 
 Astrophys.\ J.\ {\bf 546}, L29 (2001).









\bibitem{other}
For a summary of some other GRB luminosity indicators and their potential use to cosmology, see:~  E.~W.~Liang and B.~Zhang,
  arXiv:astro-ph/0504404.







\bibitem{snapdist}
  G.~Aldering {\it et al.}  [SNAP Collaboration],
  arXiv:astro-ph/0405232.





\bibitem{hubble}
  B.~E.~Schaefer,
  Astrophys.\ J.\  {\bf 583}, L67 (2003)
  [arXiv:astro-ph/0212445].




\bibitem{dust}
  A.~N.~Aguirre,
  Astrophys.\ J.\  {\bf 525}, 583 (1999)
  [arXiv:astro-ph/9904319];
  A.~Goobar, L.~Bergstrom and E.~Mortsell,
  arXiv:astro-ph/0201012;
  J.~T.~Simonsen and S.~Hannestad,
  Astron.\ Astrophys.\  {\bf 351}, 1 (1999)
  [arXiv:astro-ph/9909225].

\bibitem{dustproblems}
  A.~G.~Riess {\it et al.}  [Supernova Search Team Collaboration],
  Astrophys.\ J.\  {\bf 607}, 665 (2004)
  [arXiv:astro-ph/0402512];
 K.~Krisciunas, N.~C.~Hastings, K.~Loomis, R.~McMillan, A.~Rest, A.~G.~Riess and C.~Stubbs,
  arXiv:astro-ph/9912219.

\bibitem{axion}
  C.~Csaki, N.~Kaloper and J.~Terning,
  Phys.\ Rev.\ Lett.\  {\bf 88}, 161302 (2002)
  [arXiv:hep-ph/0111311];
  C.~Csaki, N.~Kaloper and J.~Terning,
  Phys.\ Lett.\ B {\bf 535}, 33 (2002)
  [arXiv:hep-ph/0112212];
 E.~Mortsell, L.~Bergstrom and A.~Goobar,
  Phys.\ Rev.\ D {\bf 66}, 047702 (2002)
  [arXiv:astro-ph/0202153];
  Y.~Grossman, S.~Roy and J.~Zupan,
  Phys.\ Lett.\ B {\bf 543}, 23 (2002)
  [arXiv:hep-ph/0204216].


\bibitem{lensing}
  R.~Kantowski, T.~Vaughan and D.~Branch,
  Astrophys.\ J.\  {\bf 447}, 35 (1995)
  [arXiv:astro-ph/9511108];
  J.~A.~Frieman,
  Comments Astrophys. {\bf 18}, 323 (1997)
  [arXiv:astro-ph/9608068];
  J.~Wambsganss, R.~y.~Cen, G.~h.~Xu and J.~P.~Ostriker,
  Astrophys.\ J.\  {\bf 475}, L81 (1997)
  [arXiv:astro-ph/9607084];
  P.~Valageas,
  Astron. Astrophys.  {\bf 354}, 767 (2000)
  [arXiv:astro-ph/9911336];
R.~B.~Metcalf,
 Mon. Not. Roy. Astron. Soc. {\bf 305}, 746 (1999); 
  Y.~Wang, D.~E.~Holz and D.~Munshi,
  Astrophys.\ J.\  {\bf 572}, L15 (2002)
  [arXiv:astro-ph/0204169];
D.~E.~Holz and E.~V.~Linder,
  Astrophys.\ J.\  {\bf 631}, 678 (2005)
  [arXiv:astro-ph/0412173];
   S.~Dodelson and A.~Vallinotto,
  arXiv:astro-ph/0511086.
  
\bibitem{other2}
  K.~Takahashi, M.~Oguri, K.~Kotake and H.~Ohno,
  arXiv:astro-ph/0305260;
 O.~Bertolami and P.~Tavares Silva,
  arXiv:astro-ph/0507192;
 T.~Di Girolamo, R.~Catena, M.~Vietri and G.~Di Sciascio,
  JCAP {\bf 04}, 008 (2005)
  [arXiv:astro-ph/0504591];
  E.~Mortsell and J.~Sollerman,
  JCAP {\bf 0506}, 009 (2005)
  [arXiv:astro-ph/0504245];
 C.~Firmani, G.~Ghisellini, G.~Ghirlanda and V.~Avila-Reese,
  Mon.\ Not.\ Roy.\ Astron.\ Soc.\  {\bf 360}, L1 (2005)
  [arXiv:astro-ph/0501395];
 A.~S.~Friedman and J.~S.~Bloom,
  Astrophys.\ J.\  {\bf 627}, 1 (2005)
  [arXiv:astro-ph/0408413];
  Z.~G.~Dai, E.~W.~Liang and D.~Xu,
  Astrophys.\ J.\  {\bf 612}, L101 (2004)
  [arXiv:astro-ph/0407497].

\bibitem{book}
For some examples, see Chapter 11 of S.~Dodelson, {\it Modern Cosmology} (Amsterdam, Academic Press, 2003).
      
    \bibitem{wang}
  Y.~Wang and M.~Tegmark,
  Phys.\ Rev.\ D {\bf 71}, 103513 (2005)
  [arXiv:astro-ph/0501351].

 
\bibitem{quigg1}
 G.~Barenboim, O.~Mena and C.~Quigg,
  Phys.\ Rev.\ D {\bf 71}, 063533 (2005)
  [arXiv:astro-ph/0412010].


\bibitem{quigg2}
  G.~Barenboim, O.~Mena Requejo and C.~Quigg,
  arXiv:astro-ph/0510178.









\end{thebibliography}
\end{document}